# MULTILAYER REPRESENTATION AND MULTISCALE ANALYSIS ON DATA NETWORKS

Luz Angela Aristizábal Q.[1] and Nicolás Toro G.[2]

[1]Department of Computation, Faculty of Management, Universidad Nacional de Colombia.

[2]Department of Electrical and Electronic Engineering, Universidad Nacional de Colombia.

## ABSTRACT

*The constant increase in the complexity of data networks motivates the search for strategies that make it possible to reduce current monitoring times. This paper shows the way in which multilayer network representation and the application of multiscale analysis techniques, as applied to software-defined networks, allows for the visualization of anomalies from "coarse views of the network topology". This implies the analysis of fewer data, and consequently the reduction of the time that a process takes to monitor the network. The fact that software-defined networks allow for the obtention of a global view of network behavior facilitates detail recovery from affected zones detected in monitoring processes. The method is evaluated by calculating the reduction factor of nodes, checked during anomaly detection, with respect to the total number of nodes in the network.*

## 1. INTRODUCTION

Complexity generated by the high number of devices connected to a data network [1] has been approached using strategies which tend to improve routing processes, so as to improve performance with bio-inspired techniques, based on artificial intelligence [2] and reduce monitoring times via their implementation in distributed forms [3]. This increase in the number of devices motivates not only the search for techniques that allow for reductions in node numbers analyzed by network monitors, but also the use of methods that allow for visualization of traffic information, in correlation with network topology [4].

The goal of the present document is to show the way in which the application of the multiscale transformation technique, in one network, facilitates the visual detection of anomalies and the reduction of monitoring time. The proposed strategy makes use of two current technologies: Graph Signal Processing (GSP) [15]. and Software Defined Networks (SDN).

Graph Signal Processing (GSP) is a new area of study in digital signal processing that provides conceptual and practical tools with which to model complex networks and graphically show their evolution The advantage of applying GSP to data network analysis is the possibility of relating network topology with its behavior over time [5]

Multiscale transformation methods can reveal singularities or irregular network behavior to different resolution levels. It is possible to reduce the dimensionality of analyzed data, via "coarse views of the network", which implies the analysis of lower node numbers, and consequently, a reduction of the time that network monitoring processes require [6]. This work uses previous results of the multiscale transformation applied in the reduction of monitoring registers where even though the number of data is decreased, the details of abrupt changes are preserved [7]

With the emergence of SDN in 2008, a new prospect for the implementation of network monitors was visualized [14]. In this operation model, the switches are connected to a centralized controller, each switch, at regular intervals, provides statistic information to its controller. This information is associated with the data flows circulating through its ports, which allows the subsequent generation of traffic signals that feed the network graph representation model [8][9][10].

The main contribution of this study is to show how a formal graphical representation of a network, based on multilayers and the GSP theory, facilitates the visualization of anomalous activity in an SDN network, and how, by applying multiscale analysis, the number of nodes checked in the process of anomaly detection is reduced, which accelerates the process. To illustrate the effectiveness of the method, the results of its application, oriented toward the detection of congestion, are shown.

This document includes four sections : the first defines the formal multilayer representation of the network, the second explains multiscale analysis, the third presents the application of the method, and the fourth results analysis.

## 2. NETWORK REPRESENTATION

Herein, the interest was to show a graphic visualization of the solution, guided by the premise that, "an image is worth a thousand words", and multilayer network representation was chosen. A multilayer network is formed by several layers, each with its nodes or vertices and links. The connection between layers is formed by an "intra-layer" link set. This section elucidates its formal notation [11][13].

### 2.1 Representation.

A network multilayer is defined as a pair, $\mathcal{M} = (\mathcal{G}, \mathcal{C})$, where $\mathcal{G} = \{G_\alpha : \alpha \in \{1, ...., M\}\}$ is a graph family, $G_\alpha = (X_\alpha, E_\alpha)$ is called a layer of $\mathcal{M}$, and

$$\mathcal{C} = \{E_\alpha \subseteq X_\alpha \times X_\beta \; ; \; \alpha, \beta \in \{1,...,M\}, \alpha \neq \beta\} \tag{1}$$

crosses link between the nodes of different layers, $G_\alpha$ y $G_\beta$ with $\alpha \neq \beta$. Elements of $\mathcal{C}$ are called elements of crossing layers, and elements of each $E$ are called intra-layer connections of $\mathcal{M}$, in contrast to the elements of each $E_{\alpha\beta}$ ($\alpha \neq \beta$), which are called interlayer connections. The node set of layer $G_\alpha$ is noted as $X_\alpha = \{x_1^\alpha, ...., x_{N_\alpha}^\alpha\}$ and the adjacent matrix for each layer $G_\alpha$ is $A^{[\alpha]} = (a_{ij}^\alpha) \in \mathbb{R}^{N_\alpha \times N_\alpha}$ .

$$\text{Where: } a_{ij}^\alpha = \begin{cases} 1 & if \; (x_i^\alpha, x_j^\alpha) \in E_\alpha \\ 0 & another \; case \end{cases} \tag{2}$$

For $1 \leq i, j \leq N_\alpha$ y $1 \leq \alpha \leq M$ the adjacent matrix interlayer of $E_{\alpha\beta}$ is the matrix

$$A^{[\alpha,\beta]} = \left(a_{ij}^{\alpha\beta}\right) \in \mathbb{R}^{N_\alpha \times N_\beta} \text{ give for:}$$

$$a_{ij}^{\alpha\beta} = \begin{cases} 1 & if \; \left(x_i^\alpha, x_j^\beta\right) \in E_{\alpha\beta} \\ 0 & another\; case \end{cases} \quad (3)$$

The projection network of $\mathcal{M}$ is the graph projection $(\mathcal{M}) = (X_\mathcal{M}, E_\mathcal{M})$ where

$$X_\mathcal{M} = (\cup_{\alpha=1}^M E_\alpha) \cup \left( \bigcup_{\substack{\alpha,\beta=1 \\ \alpha \neq \beta}}^M E_{\alpha\beta} \right). \quad (4)$$

In this case, the two layer $\mathcal{G}_\alpha$, $\alpha \in \{1,2\}$ was considered. The first layer, $G_1$, is formed by the nodes of the data network, $X_1$, and its respective "intra-layer" connection, $E_1$. The second layer, $G_2$, is formed by the nodes of the monitor network, $X_2$, with its "intra--layer" connection, $E_2$. The two layers are related through the connections set between data network nodes and monitor network nodes, or the cross-layer elements, $X_1 \times X_2$. Figure 1. illustrates this representation.

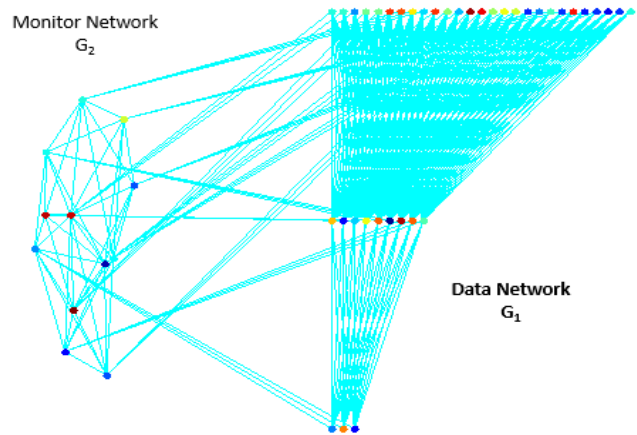

Figure 1. Monitor network and data network

## 2.2 Data in network node

The underlying network data is obtained from statistical information provided by the switches of the Software Defined Network (SDN) [12][17]. The controller periodically requests statistical information from the switch by sending the *statistic request* message, the switch provides information to the controller through the *statistics reply* message, which contains number of packets and bytes transmitted and received as well as information related to transmission errors. With this information, it is possible to form those data sequences that will constitute network traffic data for a specific time interval

The data network $G_1$ stores information in the nodes, $X_1 = \{x_1^1, \ldots, x_{N_1}^1\}$, each $x_i^1$ contains the bytes received by the openflow switch. The monitor network $G_2$ has the nodes $X_2 = \{x_1^2, \ldots, x_{N_2}^2\}$.

Where $x_i^2 = \sum_{k \in \mathbb{S}} x_k^1$ (5)

With $\mathbb{S}$, the $G_1$ node set is sensed by node $i$ from layer $G_2$. Each layer, with its adjacent matrix, is defined as (5).

Graphically, each value of $X_1$ has a color that represents the number of bytes received from the open-flow switch. Figure 2 shows this correspondence.

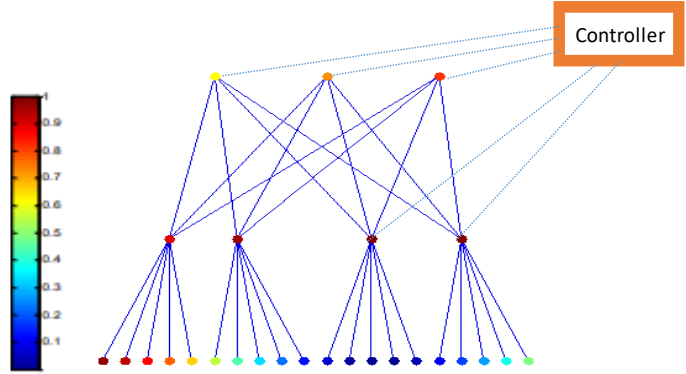

Figure 2. Correspondence between network traffic and color

This figure shows all openflow switches providing traffic information to the controller and performs the data forwarding actions that the controller determines according to the policies that have been defined by the network administrator. A controller is a piece of software that determines what actions the switch will take on the incoming flow. The communication between switches and controllers is stablished by using the protocol openflow [7]. An Openflow switch is made up of data flow tables. Each entry in the table specifies the identifier of the incoming flow, the action to be taken on the incoming flow, and a field that includes statistical information.

The red color of the figure corresponds high traffic, while dark blue corresponds little traffic in layer $G_1$. Lower-level nodes are terminal devices such as: servers, computers, printers. Nodes from the two upper levels are interconnection devices such as: openflow switches.

## 3. MULTISCALE ON SOFTWARE DEFINED NETWORKS

Multiscale methods allow for the revelation of singularities or irregular patterns for different resolution levels. At the same time, via the reduction of dimensionality in each level, it reduces task processing complexity [19].

The wavelet coefficients at scale **s**, centered on vertex **n** is defined as [20]:

$$W_x(n,s) = \sum_{j=0}^{N-1} g(s£_j)\hat{x}(j)\lambda_j(n) \qquad (6)$$

Where:

  x: graph signal.
  $\hat{x}$: graph Fourier Coefficients
  £: Laplacian matrix
  λ: eigenvector of Laplacian matrix
  N: nodes number of graph x
  g: kernel function that behaves as a band-filter

In order to reduce the nodes, the multiscale wavelet transform is applied to the graph, as input the transform take graph $x^j$ and graph Laplacian matrix $£^j$, with the j resolution level. The output for each level, $x^{j+1}$ is a rough approximation of $x^j$, which is obtained via the application of a low-band filter and downsampling. Figure 3 illustrates this process

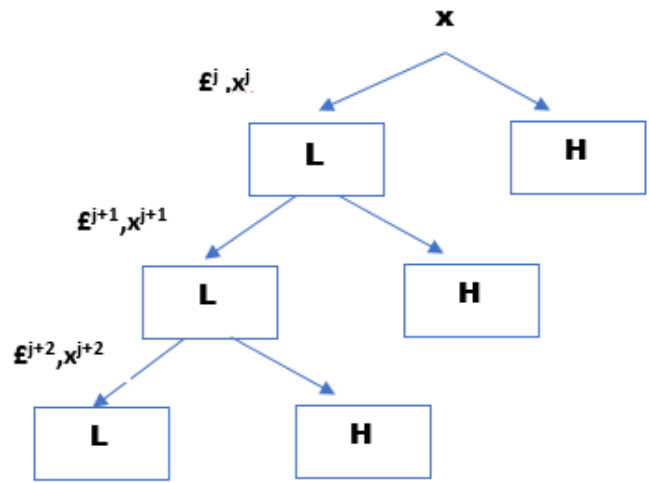

Figure 3. Pyramid transformation

The multiscale analysis begins with a network, such as that shown in Figure 4. Each node in the $G_2$ layer contains the quantified traffic of the monitored nodes, located in layer $G_1$.

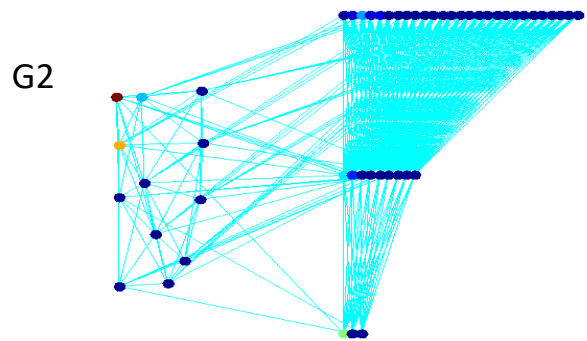

Figure 4. The monitor network with node values computed by the data network (formed by two)

When, in the $G_1$ layer, singularity is present, this is sensed by the $G_2$ layer, and visualized with the analysis of its spectral behavior. See Figure 5.

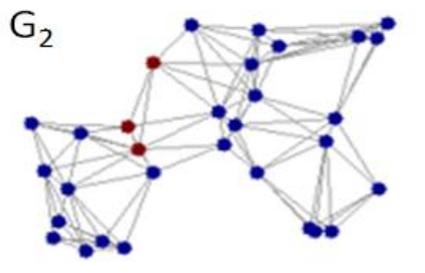

Figure 5. Monitor network

Monitor networks present a singularity or congested region, visualized in the layer's red nodes. Based on the hypothesis that, "A data network under normal conditions presents little variation (low frequencies), and that a relatively abrupt or atypical change in the behavior of the node would result in the appearance of high frequencies", the spectral analysis was carried out, and its results are shown in Figure 6.

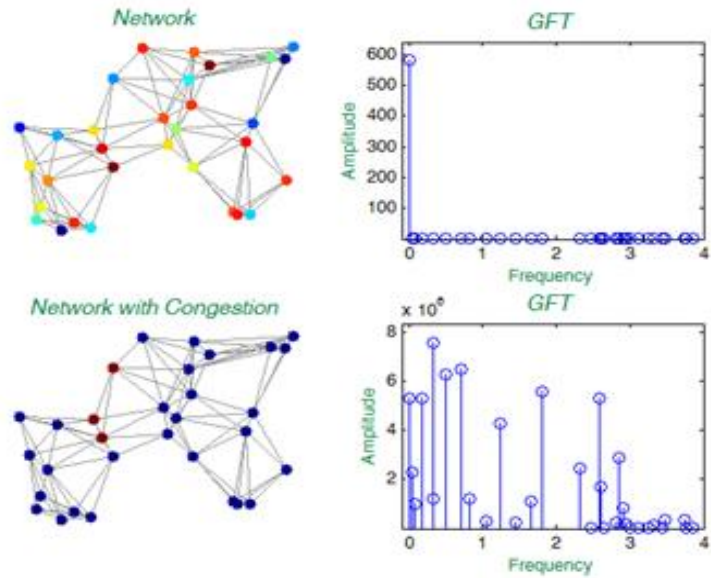

Figure 6. Spectral Analysis

This figure shows, in the upper part, the spectrum of a network with normal traffic, with its low frequency components, while the congested network has a spectrum with high frequency components. Then, the multiscale process is applied to the monitor network with the generation of $L$ resolution levels. Figure 7, shows resolution levels for the network of Figure 4, with $L=3$. Level$_{(i+1)}$ is a rougher view of previous level$_{(i)}$. This is the result of applying the wavelet transformation to $X_2 = \{x_1^2, \ldots, x_{N_2}^2\}$.

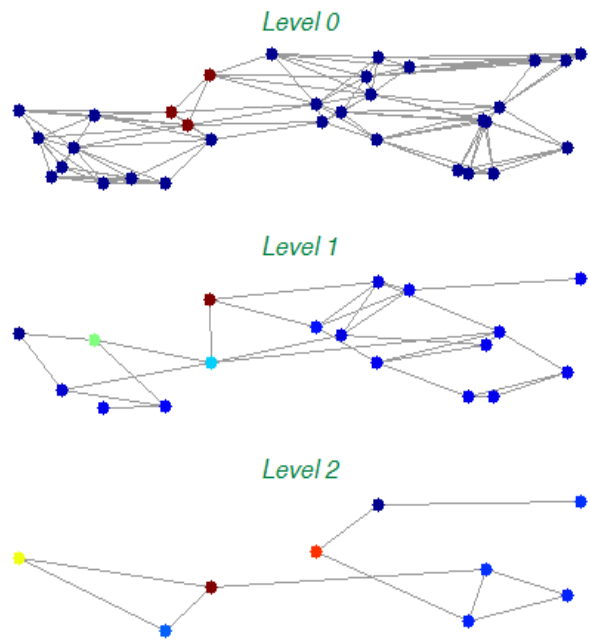

Figure 7. Three resolution levels

The implementation of multiresolution decomposition is achieved with graphic wavelet transform, which is made up of a bank of filters: one low pass filter (LP) and high pass (HP), whose coefficients are determined by the base wavelet [16][18].

Considering the results obtained with the spectral analysis shown in Figure 6, the HP filter output was selected to detect anomalies (in this case, congestion detection). This occurred given that the spectrum of each resolution level presents high frequencies, and this spurs the search for the anomalous regions. Spectral analysis of resolution levels is shown in Figure 8.

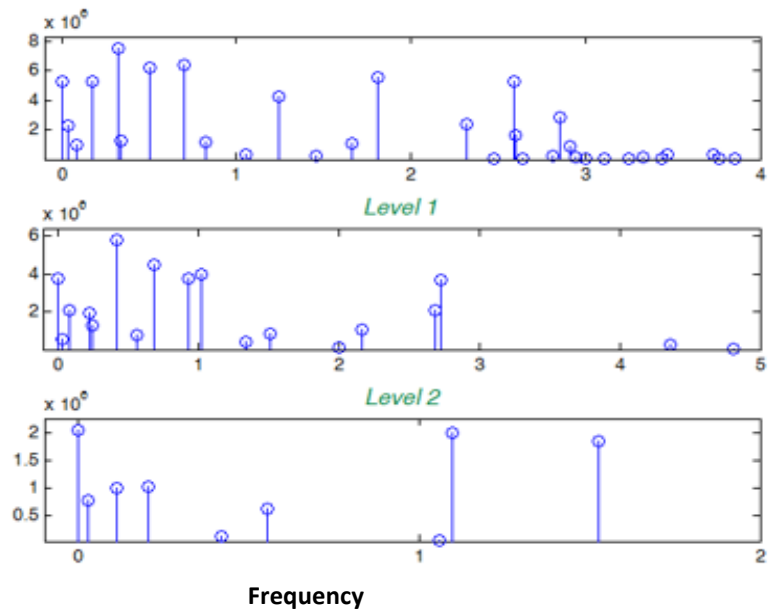

Figure 8. Spectral analysis of resolution levels

## 4. APPLICATION OF METHOD

The traffic used to perform the test was taken from simulations of software-defined networks, implemented in a Mininet simulation environment, the controller implemented in Python. Initially, the statistics sent by the SDN switches to the controller were used as a response to the "*statistic request*" message. With this information, a structure was formed, which contained those bytes transmitted and received by each of the devices connected to the network switches. The aim was to form a graph that contained the information received by the controller.
The algorithm is shown in the figures below:

1. Generation of data network ($G_1$) and instantiation of a multilayer network with $G_1$ and $G_2$. The data network has a structure of datacenter with only two layers of openflow switches on a topology Leaf-Spine (with connection between all nodes).

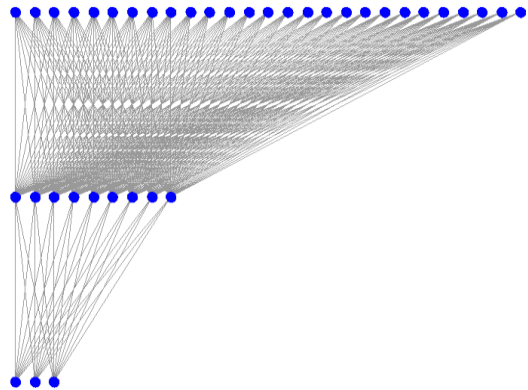

Figure 9. Data center topology and multilayer network

2. Design of base filter of wavelet transform. Figure 10 shows the frequency response.

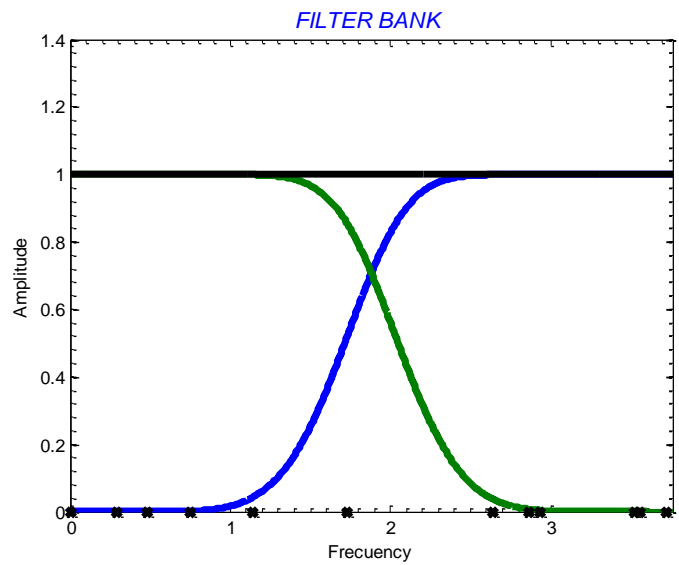

Figure 10. Filter bank used in wavelet transform

3. With the monitor layer $G_2$, Fig 11, the wavelet transform was applied to obtain resolution levels

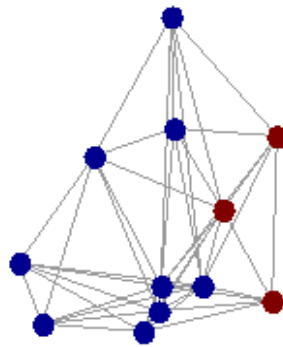

Figure 11. $G_2$ layer with congestion

Figure 12, shows singularity by level, in this case, congestion.

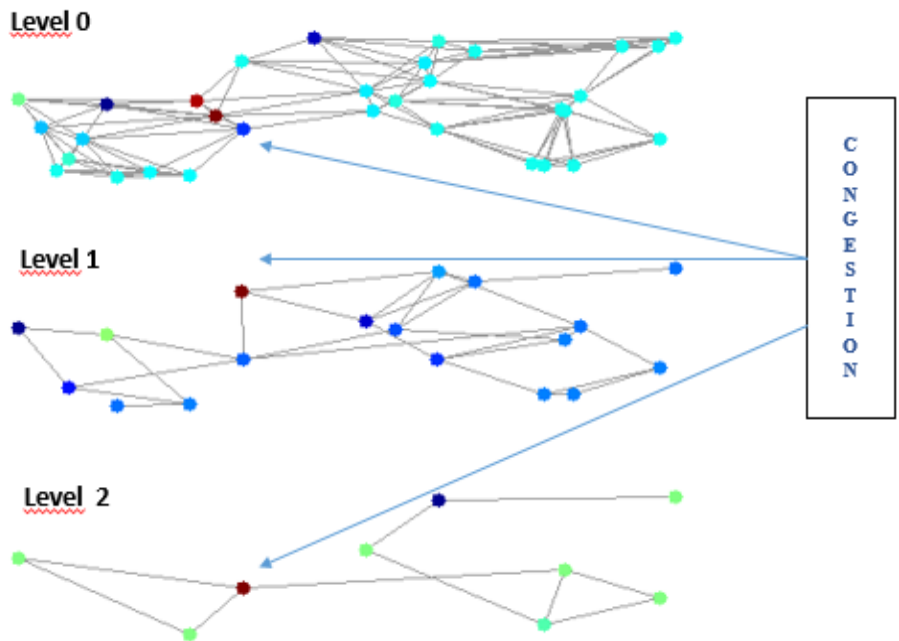

Figure 12. Congestion in each layer

4. Spectral analysis and thresholding on the rougher level of the multiscale decomposition. This analysis is performed for identified possible congestion zones. Figure 13 shows one congestion zone (red dot).

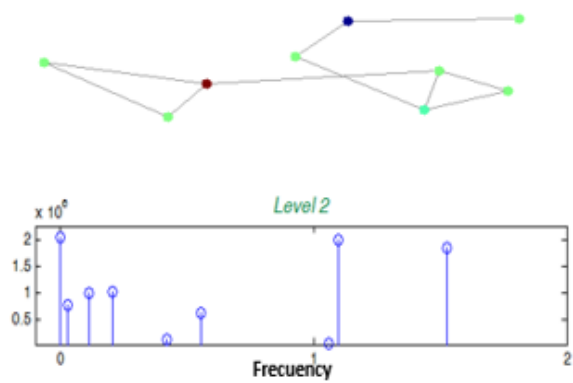

Figure 13. Last level of decomposition

5. If the spectrum of the rougher level has high frequencies, or singularity zones, it begin a search process of those nodes with highest color intensity and its references to nodes in the upper layer. This is shown in Figure 14. The node with highest intensity, in red, indicates the presence of congestion in the network. With this information, the algorithm looks for the correspondence location of this node, in the upper level. The search is repeated until it reaches the highest level of decomposition, or the initial monitoring network,

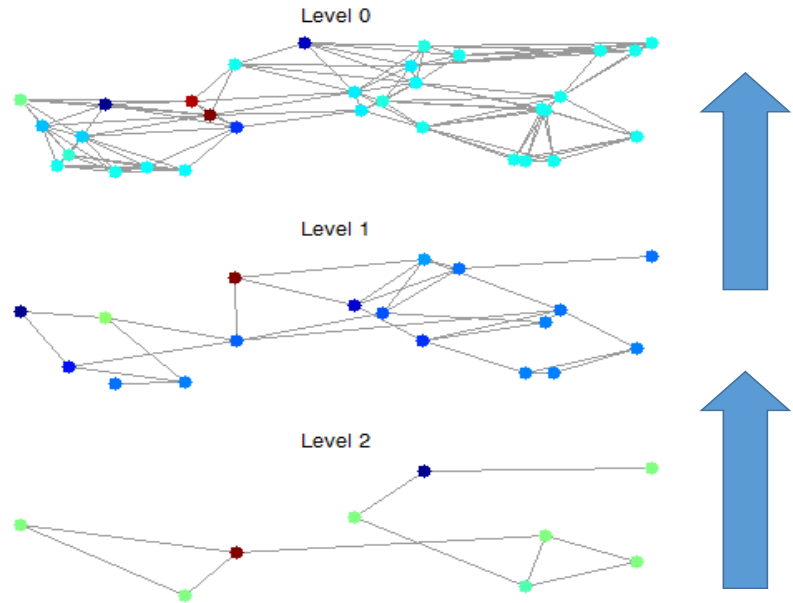

Figure 14. Congestion zone detection beginning on the roughest level

## 5. RESULT ANALYSIS

Results are evaluated in light of two aspects: the ratio of nodes reduction in the visual detection of anomalies and the measure of computational complexity in the multiresolution analysis method.

1. The reduction of nodes number is implemented in two stages:
    a. The first reduction occurs when the monitor $G_2$ layer, which supervises the behavior of the $G_1$ layer nodes, the initial network, is generated.
       In other words:
       $$|G_2| = \frac{|G_1|}{\log_2 |G_1|} \quad (7)$$
       Where:
       $|G_2|$: Node numbers in the $G_2$ layer
       $|G_1|$: Node numbers in the $G_1$ layer

    Each node in the $G_2$ layer monitors $\log_2 |G_1|$ nodes in the $G_1$ layer. The following figure shows the relationship between the node numbers in these two layers.

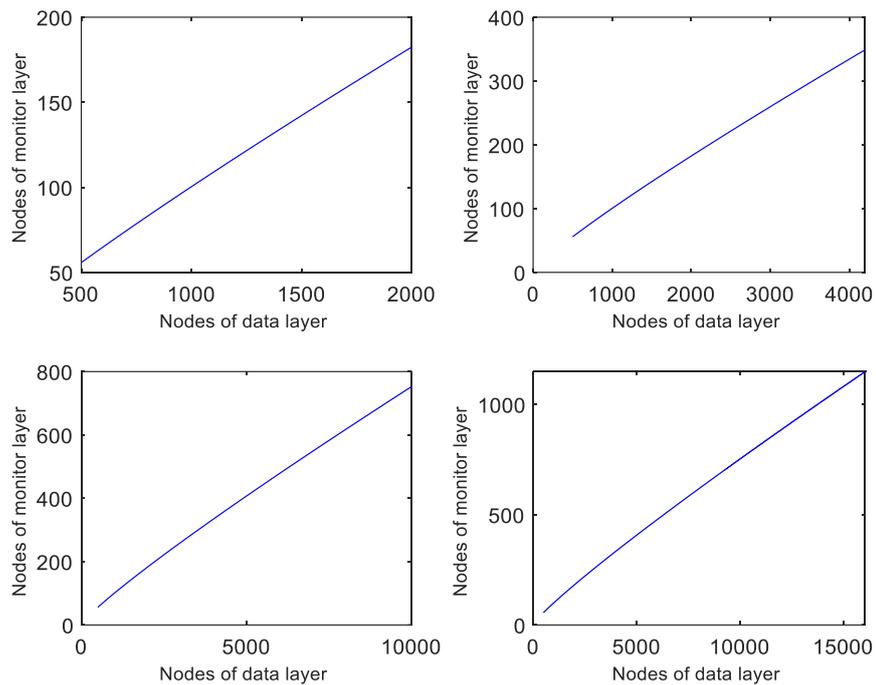

Figure 15. Relationships between nodes from the $G_1$ layer (x axis) and $G_2$ layer (y axis)

These graphs reflect a substantial reduction of nodes, by allowing several nodes from the initial data network, $G_1$ layer to be monitored by a single node from the $G_2$ layer. It is relevant to note that a single node from the $G_2$ layer monitors a considerably low number of nodes from the $G_1$ layer. For example, for a data network whose number of nodes is in the order of 16,000, the $G_2$ layer will have 1,148 nodes, each of these monitoring only 14 nodes. Thus, it is desirable to avoid the camouflaging of any abnormality in network.

b. The second reduction is implemented in the multiscale analysis. The node numbers of a resolution level depend on the node numbers in the upper level. Each resolution level with N nodes is reduced to $N/2$ nodes for the next level. See Figure 12

Thus, if we begin with a network with 16,000 nodes, the first reduction will deliver 1,148 nodes to multiresolution analysis. If this is applied twice, the process will end with 287 nodes, or if applied once, it will finish with 574 nodes .The network formed with these nodes will be the entry network with which to process anomaly detection.

2. The efficiency of computational implementation: The analysis of this point should answering the follow question: What is less expensive, the analysis of N nodes, been N large, how is in datacenters or apply multiresolution analysis to reduce nodes number, prior to an anomaly detection process?

Considerations:
- The method outlined in this article allows the network administrator or operator, or monitoring application to be aware of the presence of anomalies. This occurs via graphical Fourier analysis. Said analysis is performed after initial nodes reduction.

- In the case of abnormalities, if the number of nodes is quite large, the multiresolution method is applied to detect its location in the network. The implementation of graph wavelet transformation, via polynomial approximation, has a computational complexity of:
  O(M|E| + MN (J + 1))  in [18]

  Where:
  M: is the degree of polynomial approximations for each of the scaled wavelet kernels.
  |E|: is the total number of non-zero edges in the graph, with sparse matrix representation.
  J: is the number of levels in the decomposition process.
  N: is the total number of nodes.

The following graphs show the relationship between the computational cost of multiresolution analysis and the variation of polynomial degrees of wavelet transformation

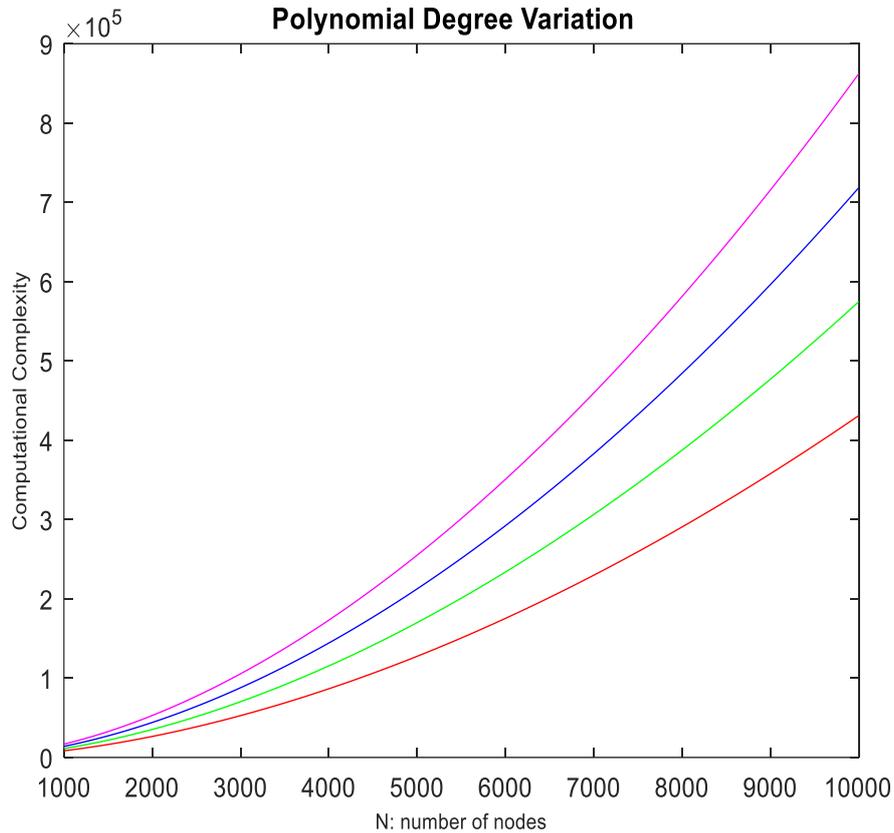

Figure 16. The computational complexity of multiresolution analysis. Degree=3 (red), 4 (green), 5 (blue), 6 (magenta)

The above graph was calculated for $|E| = |G_2|*|G_2 - 1|/2$, the worst case scenario, that is, where the nodes of the monitor network form a mesh topology.

> In accordance with the tests, an adequate resolution was obtained when the transformation was calculated using a polynomial of degree four, in the figure 16, the green curve.

Considering this result, the answer to question formulated in the initial part of this item will depend of the nodes number of the network. Visually for the network manager is more simple checking which is the region that presents difficulties if the nodes number is small. The network manager could up from an low level to upper level with only a complexity of O(1), as is explained in the Figure 14. The computational cost not high either, if is realized with the polynomial method.

For a nodes number small, the detection with Graph Fourier Transform is sufficient.

## 6. CONCLUSIONS

The added value of the multi-layer representation permitted two simplifications in the detection of congested areas. The first of these was related to the reduction of nodes evaluated, considering only nodes from the network of the upper layer, or sensor network. The second focused on a decrease achieved when congestion analysis initially takes the topologies with the lowest number of nodes, obtained from the multiscale analysis, or nodes of the rougher level.

The implementation of multiscale analysis on a software defined multilayer network presented the following advantages: the monitor layer was virtual, as was its topology and all processes implemented on it were carried out via the controller. Physical network devices only contributed the statistical information supplied to the controller.
It is necessary to approach this implementation in a distributed way, implementing the monitor network via several controllers. This would improve response times for a network with a large number of nodes

## ACKNOWLEDGEMENT

The research paper corresponds to "Programa reconstrucción del tejido social en zonas de posconflicto en Colombia, proyecto Modelo ecosistémico de mejoramiento rural y construcción de paz: instalación de capacidades locales," and was financed by the "Fondo Nacional de Financiamiento para la Ciencia, la Tecnología, y la Innovación, Fondo Francisco José de Caldas".

**Authors**

**Luz A. Aristizábal Q.** is a professor in the Department of Computing in the Faculty of Management at the Universidad Nacional de Colombia. She received her MEng in Physical Instrumentation from the Technological University of Pereira in 2009, her degree in Data Networks Specialization from Valle University in 1991, and her degree in Engineering Systems from Autónoma University in 1989. Her research focuses on aspects of computer and data networks, including the network simulators, signal processing and network paradigms

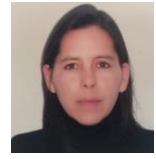

**Nicolás Toro G**. is a professor in the Department of Electrics, Electronics and Computing. He received his PhD in Engineering-Automation and SB in Electrical Engineering from the Universidad Nacional de Colombia in 2013 and 1983 respectively, and his MEng degrees in Automated production systems from the Technological University of Pereira in 1992. His research focuses on many aspects of industrial automation, including the design, measurement, and analysis of networks

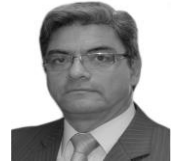